\input harvmac.tex
\noblackbox

\def\ie{{\it i.e.}}

\def\eighth{{\textstyle{1 \over 8}}}
\def\ZZ{ {\bf  Z} }
\def\e{{\rm e}}
\def\Gmnk{ G(m,n,k) }
\def\Ghmnk{ {\hat G}(m,n,k) }
\def\Gknm{ G(k,n,m) }
\def\SU{{\rm SU}}
\def\SUNK{ \SU(N)_K }
\def\SUKN{ \SU(K)_N }
\def\SO{{\rm SO}}
\def\U{{\rm U(1)}}
\def\Gh{ {\hat G}}
\def\Sh{ {\hat S}}
\def\Th{ {\hat T}}
\def\Nh{ {\hat N}}   
\def\Kh{ {\hat K}}   
\def\mh{ {\hat m}}   
\def\nh{ {\hat n}}   
\def\kh{ {\hat k}}   
\def\hh{ {\hat h}}   
\def\ch{ {\hat c}}   
\def\mod{{\rm ~~~mod~}}
\def\La{   \Lambda           }
\def\lla{   \lambda           }
\def\lam{   \lambda           }
\def\la{   {\lambda_0}           }
\def\lo{   {\lambda_1}           }
\def\lt{   {\lambda_2}           }
\def\Lah{   {\hat{\La}}      }
\def\llah{   {\hat{\lla}}      }
\def\lamh{   {\hat{\lambda}}      }
\def\lah{   {\lamh_0}         }
\def\loh{   {\lamh_1}         }
\def\lth{   {\lamh_2}         }
\def\Lat{  {\tilde{\La}}     }
\def\llat{  {\tilde{\lla}}     }
\def\lamt{  {\tilde{\lambda}}     }
\def\lat{  {\lamt_0}          }
\def\lot{  {\lamt_1}          }
\def\lamb{  {\bar{\lambda}}       }
\def\llab{  {\bar{\lambda}}       }
\def\lab{  {\lamb_0}          }
\def\lob{  {\lamb_1}          }
\def\ltb{  {\lamb_2}          }
\def\Lath{ {\hat{\Lat}}      }
\def\lamth{ {\hat{\lamt}}      }
\def\lath{ {\lamth_0}         } 
\def\loth{ {\lamth_1}         } 
\def\lamht{ {\tilde{\lamh}}    }
\def\laht{ {\lamht_0}         }
\def\loht{ {\lamht_1}         }
\def\lthb{ {{\bar{\lamh}}_2}  }
\def\Lap{  {\La^\prime}      }
\def\lamp{  {\lambda^\prime}      }
\def\lap{  {\lambda_0^\prime}      }
\def\lop{  {\lambda_1^\prime}    }
\def\ltp{  {\lambda_2^\prime}    }
\def\Lahp{  {\Lah^\prime}      }
\def\lahp{ {\lamh_0^\prime}     }
\def\lohp{ {\lamh_1^\prime}   }
\def\lthp{ {\lamh_2^\prime}   }
\def\Latp{ {\Lat^\prime}     }
\def\lamtp{ {\lamt^\prime}     }
\def\latp{ {\lamt_0^\prime}     }
\def\lotp{ {\lamt_1^\prime}   }
\def\ltbp{ {\lamb_2^\prime}   }
\def\lahtp{{\lamht_0^\prime}    }
\def\lohtp{{\lamht_1^\prime}   }
\def\lthbp{{{\bar{\lamh}}_2^\prime}   }
\def\eps{  \epsilon          }
\def\ta{ {\tilde{a}}          }
\def\qp{   {q^\prime }      }
\def\qh{   {\hat{q}}        }
\def\qhp{  {\qh^\prime}      }
\def\qt{   {\tilde{q}}      }
\def\qtp{  {\qt^\prime}      }
\def\qth{  {\hat{\qt}}       }
\def\qthp{ {\qth^\prime}     }
\def\Sig{  J            }
\def\Pie{  f               }

\def\xt{  {\tilde{x}}        }
\def\xp{  {x^\prime}         }
\def\mue{  {\mu_0}           }
\def\mo{   {\mu_1}           }
\def\mt{   {\mu_2}           }
\def\Qt{   {\tilde{Q}}      }
\def\piep{ {\pi^\prime}      }
\def\epsp{ {\eps^\prime}     }
\def\ip{   {i^\prime}        }
\def\alt{  {\tilde{a}  }   }
\def\bet{  {\tilde{b}  }   }
\def\epst{ {\tilde{\eps} }   }
\def\uup{   {u^\prime }    }
\def\vp{   {v^\prime}     }
\def\exp{{\rm exp}}
\def\si{\sigma}

\def\undertext#1{\vtop{\hbox{#1}\kern 1pt \hrule}}
\lref\BH{K. Bardakci and  M. Halpern, 
             {\sl  Phys. Rev.} {\bf D3} (1971) 2493;\hfil\break
             M. Halpern, 
             {\sl  Phys. Rev.} {\bf D4} (1971) 2398}
\lref\GKO{P. Goddard, A. Kent, and D. Olive,
             {\sl  Phys. Lett.} {\bf B152} (1985) 88;\hfil\break
          P. Goddard, A. Kent, and D. Olive,
             {\sl  Commun. Math. Phys.} {\bf 103} (1986) 105}
\lref\Kazama{Y. Kazama and H. Suzuki,
             {\sl  Nucl. Phys.} {\bf B321} (1989) 232}
\lref\Moore{G. Moore and N. Seiberg,
             {\sl Phys. Lett.} {\bf B220} (1989) 422}
\lref\GepnerPL{D. Gepner,
             {\sl  Phys. Lett.} {\bf B222} (1989) 207}
\lref\Lerche{W. Lerche, C. Vafa, and N. P. Warner,
             {\sl  Nucl. Phys.} {\bf B324} (1989) 427}
\lref\SYCoset{A. N. Schellekens and S. Yankielowicz, 
             {\sl  Nucl. Phys.} {\bf B334} (1990) 67}
\lref\Schellekens{A. N. Schellekens, 
             {\sl  Nucl. Phys.} {\bf B366} (1991) 27}
\lref\FSS{J. Fuchs, A. N.  Schellekens, and C. Schweigert,
              {\sl Nucl. Phys.} {\bf B461} (1996) 371}
\lref\GepnerNP{D. Gepner,
             {\sl  Nucl. Phys.} {\bf B322} (1989) 65}
\lref\BHT{M. Blau, F. Hussain, and G. Thompson, 
             {\sl  Nucl. Phys.} {\bf B488} (1997) 541;\hfil\break
             {\sl  Nucl. Phys.} {\bf B488} (1997) 599}
\lref\Verlinde{E. Verlinde,
             {\sl  Nucl. Phys.} {\bf B300} [FS22] (1988) 360}
\lref\LevelRank{S. Naculich and H. Schnitzer,
             {\sl   Phys. Lett.} {\bf B244} (1990) 235;\hfil\break
                S. Naculich and H. Schnitzer,
             {\sl  Nucl. Phys.} {\bf B347} (1990) 687;\hfil\break
              S. Naculich, H. Riggs, and H. Schnitzer,
             {\sl   Phys. Lett.} {\bf B246} (1990) 417 }
\lref\Kuniba{A. Kuniba and T. Nakanishi, 
              in {\it Modern Quantum Field Theory},
               S. Das. {\it et. al.}, eds.
               (World Scientific, Singapore, 1991)}
\lref\ABI{D. Altsch\"uler, M. Bauer, and C. Itzykson,
             {\sl Commun. Math. Phys.} {\bf 132} (1990) 349}
\lref\Mlawer{E. Mlawer, S. Naculich, H. Riggs, and H. Schnitzer,
             {\sl  Nucl. Phys.} {\bf B352} (1991) 863}
\lref\Nakanishi{T. Nakanishi and A. Tsuchiya,
             {\sl Commun. Math. Phys.} {\bf 144} (1992) 351}
\lref\Altschuler{D. Altsch\"uler,
             {\sl  Nucl. Phys.} {\bf B313} (1989) 293}
\lref\ABS{D. Altsch\"uler, M. Bauer, and H. Saleur,
             {\sl  Jour. Phys.} {\bf A23} (1990) L789}
\lref\Frenkel{I. Frenkel, in 
           \it Lie Algebras and Related Topics,
           \rm Lecture Notes in Mathematics, no. 933, 
             D. Winter, ed. (Springer-Verlag, Berlin, 1982)}
\lref\FSAnn{ J. Fuchs and C. Schweigert,
             {\sl  Ann. Phys.} {\bf 234} (1994) 102}
\lref\FSNP{  J. Fuchs and C. Schweigert,
             {\sl  Nucl. Phys.} {\bf B411} (1994) 181}
\lref\Kac{   V. Kac and M. Wakimoto, 
              \sl Adv. Math. \bf 70 \rm (1988) 156}
\lref\Fuchs{J. Fuchs and D. Gepner, 
             {\sl  Nucl. Phys.} {\bf B294} (1987) 30}
\lref\SYExt{A. N. Schellekens and S. Yankielowicz, 
             {\sl  Nucl. Phys.} {\bf B327} (1989) 673;\hfil\break
	    K. Intriligator,
             {\sl  Nucl. Phys.} {\bf B332} (1990) 541}
\lref\Zuber{J.-B. Zuber,
             {\sl  Commun. Math. Phys.} {\bf 179} (1996) 265;\hfil\break
		S. Gusein-Zade and A. Varchenko,
		hep-th/9610058}
\lref\SYReview{A. N. Schellekens and S. Yankielowicz, 
             {\sl  Int. Jour. Mod. Phys.} {\bf A5} (1990) 2903}
\lref\SYApp{Appendix A of ref.~\SYCoset.}
\Title{\vbox{
\hbox{\tt hep-th/9705149}
\hbox{HUTP-97/A019}
\hbox{BOW-PH-108}
\hbox{BRX-TH-413}
}} 
{\vbox{\centerline{Superconformal Coset Equivalence}
\medskip
\centerline{from Level-Rank Duality}}}
\centerline{ 
Stephen G. Naculich\footnote{$^\dagger$}
{Supported in part by the DOE under grant DE-FG02-92ER40706}
{}\footnote{$^*$}
{naculich@polar.bowdoin.edu}
}
\smallskip
\centerline{\it Lyman Laboratory of Physics, Harvard
University, Cambridge, MA 02138}
\smallskip
\centerline{\it Department of Physics, Bowdoin College,
Brunswick, ME  04011}
\smallskip
\centerline{Howard J. Schnitzer${}^\dagger${}\footnote{$^{**}$}
{schnitzer@binah.cc.brandeis.edu}
}
\smallskip\centerline{\it Department of Physics,
Brandeis University, Waltham, MA 02254}
\vskip .2in

We construct a one-to-one map between the primary fields
of the $N=2$ superconformal Kazama-Suzuki models
$G(m,n,k)$ and $G(k,n,m)$ based on complex Grassmannian cosets,
using level-rank duality of Wess-Zumino-Witten models.
We then show that conformal weights, superconformal U(1) charges,
modular transformation matrices, and fusion rules 
are preserved under this map,
providing strong evidence for the equivalence of these coset models.

\Date{May 1997} 

\newsec{Introduction}

One of the largest classes of non-trivial but still solvable
two-dimensional conformally invariant field theories 
arises from the coset construction \refs{\BH,\GKO}.
In this construction, 
the conformal algebra associated with 
the level $k$ Kac-Moody algebra of some group $G$ 
is orthogonally decomposed into 
the conformal algebra associated with some subgroup $H$
and the $G/H$ coset conformal algebra.
In a similar way, 
superconformal coset algebras can be obtained
from the orthogonal decomposition of 
the super Kac-Moody algebra associated with $G$.
Kazama and Suzuki \Kazama~showed that 
the superconformal algebra based on the coset $G/H$
possesses an extended ($N=2$) superconformal symmetry
if, for rank $G=$ rank $H$,
the coset $G/H$ is a K\"ahler manifold.
To establish that this algebra leads to 
a well-defined modular invariant conformal field theory
requires consideration of the spectrum of primary fields,
which involves issues  of 
selection rules, 
field identifications \refs{\Moore,\GepnerPL,\Lerche},
and fixed-point resolutions \refs{\SYCoset,\Schellekens,\FSS}.
In this paper, we focus on the class of Kazama-Suzuki models 
based on the complex Grassmannian manifold
$ \SU(m+n)/\left[ \SU(m) \times \SU(n) \times \U \right]$.
These superconformal coset models may be written as ordinary coset models
\eqn\DefCoset{
\Gmnk = { \SU(m+n)_k \times \SO (2mn)_1 \over
           \SU(m)_{n+k} \times \SU(n)_{m+k} \times \U_{mn(m+n)(m+n+k)}},
}    
where the $\SO(2mn)_1$ factor arises from the adjoint fermions
of the super Kac-Moody algebra.

The invariance of the central charge of the coset models \DefCoset
\eqn\Central{
c^{m,n,k} = {3 m n k \over m+n+k}
}
under any permutation of $m$, $n$, and $k$
suggests that the models themselves may be invariant \Kazama.
The invariance of the coset models 
under $m \leftrightarrow n$ is manifest from their definition,
but the further symmetry under
$k \leftrightarrow m$ 
is unexpected, 
as $k$ and $m$ play rather different roles.
Kazama and Suzuki \Kazama~showed 
that the supercurrent also respects the $k \leftrightarrow m$ symmetry,
providing further evidence for the conjecture.
Gepner \GepnerNP~demonstrated that the Landau-Ginzburg models 
corresponding to $G(m,1,k)$ and to $G(k,1,m)$ are equivalent,
and Lerche {\it et. al.} \Lerche~showed that the Poincar\'e polynomials
of $G(m,n,1)$ and $G(1,n,m)$ are identical.
See also ref.~\BHT.

In this paper, we construct an explicit one-to-one map 
between the primary fields of $\Gmnk$ and $\Gknm$ 
when $m$, $n$, and $k$ have no common divisor, 
or only a prime common divisor.
When $m$, $n$, and $k$ have greatest common divisor $p>1$, 
the model has fixed points that must be resolved 
into a multiplicity of fields to maintain modular invariance.
Schellekens \Schellekens~has shown how to do this for $p$ prime,
and for this case
we exhibit the one-to-one map between the resolved primary fields
of $\Gmnk$ and $\Gknm$.
We then demonstrate that the modular transformation matrices
$S$ and $T$ are identical in the two theories.
This  further implies the equality (modulo integers) 
of conformal weights of corresponding primary fields,
and the equality of the fusion rules via Verlinde's formula \Verlinde.
These identifications  provide nearly conclusive evidence 
that $\Gmnk$ and $\Gknm$
are equivalent conformal field theories. 
This equivalence arises largely as a consequence of the
level-rank duality 
\refs{\LevelRank,\Kuniba,\ABI,\Mlawer,\Nakanishi} 
of the constituent WZW models.

Level-rank duality has been shown to underlie equivalences
between other coset models.
Altsch\"uler \Altschuler~has shown the equivalence of the conformal
generators of various pairs of dual (non-superconformal) 
coset models, 
and the equivalence of the characters of certain 
non-unitary coset models was shown \ABS~to follow from the duality
of principally-specialized characters \Frenkel.
Most closely related to the present work is that of
Fuchs and Schweigert \FSAnn,
who used the level-rank duality 
of orthogonal and symplectic groups \Mlawer~to show the equivalence
of several pairs of $N=2$ superconformal models. 
In particular, they demonstrated the equivalence of the
Kazama-Suzuki models
\eqn\OrthoEquiv{
    { \SO(m+2)_k \times \SO (2m)_1 \over
      \SO(m)_{k+2} \times \U_{4(m+k)}   }
 =  { \SO(k+2)_m \times \SO (2k)_1 \over
      \SO(k)_{m+2} \times \U_{4(m+k)}   }
}
for $m$ and $k$ odd, and for $m$ even and $k$ odd but with a non-diagonal
modular invariant in the theory on the right hand side.
They also showed an isomorphism between several other sets of 
coset models based on non-hermitian symmetric spaces \FSNP.

This paper is organized as follows:
in section 2, we describe the Kazama-Suzuki model $\Gmnk$ in some detail. 
Section 3 reviews level-rank duality between $\SUNK$ and $\SUKN$.
In section 4, we construct the map
between primary fields of $\Gmnk$ and $\Gknm$,
and demonstrate that the conformal weights and modular
transformation matrices of corresponding fields are the same.
Section 5 describes the map between the chiral rings of
$G(m,1,k)$ and $G(k,1,m)$.
In section 6, we discuss the fixed-point resolution 
when $m$, $n$, and $k$ possess a common (prime) divisor,
and construct the one-to-one map 
between resolved primary fields.
Some concluding remarks form section 7.

\newsec{Primary fields of the complex Grassmanian coset model}

In this section, 
we describe the relevant details of the 
complex Grassmannian Kazama-Suzuki model \DefCoset.
The central charge of this model \Central~is obtained from
\eqn\CosetC{
c^{m,n,k} = c^{m+n,k} - c^{m,n+k} - c^{n,m+k} + mn - 1
}
where $c^{N,K} = K(N^2-1)/(K+N)$ denotes 
the central charge of the $\SUNK$ WZW model,
and $mn$ and $1$ are the central charges of the SO$(2mn)_1$ and U(1) models
respectively.

The characters of the coset model
in the absence of fixed-point subtleties\foot{We will
deal with fixed points in section 6.}
are given by the branching functions 
$ b^{\la,\pi}_{\lo,\lt,q} (\tau) $
in the character decomposition
\eqn\Branching{
\chi^{\la,\pi} (\tau) 
= \sum  b^{\la,\pi}_{\lo,\lt,q} (\tau)
        \chi^{\lo,\lt,q} (\tau)
}
where $\la$, $\pi$, $\lo$, $\lt$, and $q$
denote primary fields of  
$ \SU(m+n)_k $, $ \SO (2mn)_1 $, $ \SU(m)_{n+k} $, $ \SU(n)_{m+k} $, 
and $ \U_{mn(m+n)(m+n+k)} $ respectively.
Thus, primary fields of the coset model are labelled
by the multi-index
\eqn\CosetPrimaryField{
\La = \pmatrix{ \la & & \pi \cr \lo & \lt & q  }
} 
subject to the selection rules and identifications specified below.

Primary fields of the  $\SUNK$ WZW model are labelled by 
integrable representations $\lla$ of the $\SUNK$  Kac-Moody algebra, 
those with non-negative extended Dynkin indices
$\{ a_0, a_1, \ldots, a_{N-1} \}$,
where $a_0 = K - \sum_{i=1}^{N-1}  a_i $.
These representations have Young tableaux whose first row length 
$\ell_1 = \sum_{i=1}^{N-1} a_i$ 
is no greater than $K$ boxes.
Primary fields of $\SO (2N)_K$ are labelled by
integrable representations $\pi$ of $ \SO (2N)_K $,
again with non-negative extended Dynkin indices
$\{ a_0, a_1, \ldots, a_{N} \}$,
where now 
$a_0 = K - a_1 - 2 \sum_{i=2}^{N-2}  a_i - a_{N-1} - a_{N}  $.
For $\SO (2N)_1$, 
there are only four:
the singlet (1), vector ($v$), 
spinor ($s$), and conjugate spinor ($c$) representations,
which correspond to non-zero $a_0$, $a_1$, $a_{N-1}$, and $a_{N}$
respectively.
In the following, 
we will refer to the first two of these as the Neveu-Schwarz (NS) sector,
and the last two as the Ramond (R) sector,
alluding to the fermionic origin of the $\SO(2mn)_1$ factor.
$\U_L$ has $L$ primary fields, 
labelled by the integers
$q=0,1,\ldots,L-1$ mod $L$.  

\vfil\break
\noindent\undertext{Conformal weights and modular transformation matrices}

The modular transformation matrices of the coset model characters 
\eqn\Modular{\eqalign{
\chi_{\La} (-1/\tau) &= \sum_{\Lap} S_{\La\Lap} \chi_{\Lap} (\tau) \cr
\chi_{\La} (\tau+1) & = T_{\La\La} \chi_{\La} (\tau)
= \e^{  2\pi i \left( h_\La - c/24 \right) } \chi_{\La}(\tau) \cr
}}
can be inferred from those of the branching functions \Branching.
When there are no fixed points\foot{See section 6 for 
modifications when fixed points are present.},
the coset modular matrix $S$ is
\eqn\CosetS{
S^{m,n,k}_{\La \Lap} 
= {mn(m+n)} 
S_{\la \lap} S^*_{\lo \lop} S^*_{\lt \ltp}
               S_{\pi \piep} S^*_{q \qp}
}
where
$S_{\lla \lla^\prime}$ are the modular transformation matrices
of the $\SUNK$ WZW models \Kac,
\eqn\OrthoS{
S_{\pi \piep}  = \half \pmatrix{  1 & 1 & 1 & 1 \cr
                             1 & 1 & -1 & -1 \cr
                             1 & -1 & i^{-N} &  -i^{-N} \cr
                             1 & -1 & -i^{-N} & i^{-N} \cr}
{\rm ~~~~~~~~~~~~for~~SO}(2N)_1,
}
and
\eqn\UniS{
S_{q \qp} = {1 \over \sqrt{L} } \exp \left(-~{2\pi i q \qp \over L} \right)
{\rm ~~~~~~~~~~~~~~~~~~for~~U}(1)_L.
}
The factor of $mn(m+n)$ in eq.~\CosetS~results from field identification,
discussed below.
Similarly, eqs.~\Branching~and \Modular~determine 
the conformal weights of the coset primary fields modulo integers
\eqn\CosetH{
h^{m,n,k}_\La 
= h^{m+n,k}_\la - h^{m,n+k}_\lo - h^{n,m+k}_\lt + h^{2mn}_\pi - h_q
\mod \ZZ }
where
\eqn\WZWH{\eqalign{
h^{N,K}_\lla &= { C_2 (\lla) \over K+N}  
{\rm ~~~~~~~~~~~~~~~~~~~~~~~~~~~~~~~~~~~~~~~for~~SU}(N)_K \cr
h^{2N}_\pi &= (0, \half, \eighth N, \eighth N) {\rm ~~if~~}
\pi = (1, v, s, c) 
{\rm ~~~~~~for~~SO}(2N)_1 \cr
h_q &= { q^2 \over 2L } \mod \ZZ
{\rm ~~~~~~~~~~~~~~~~~~~~~~~~~~~~~~~~~for~~U}(1)_{L} \cr
}}
with $C_2 (\lla)$ the quadratic Casimir of the representation
$\lla \in \SU (N)$, here normalized to $C_2 ({\rm adjoint}) = N$.

\vfil\break
\noindent\undertext{Superconformal U(1) charges}

The $N=2$ superconformal symmetry of $ \Gmnk $
implies that the primary field $\La$
carries a superconformal U(1) charge $Q_\La$
given by \Kazama
\eqn\ScflCharge{
Q_\La = \Pie_{2mn}(\pi) - {q \over m+n+k}
\mod 2}
where
\eqn\DefSigma{
\Pie_{2N}(\pi) = (0, 1, \half N, \half N -1) {\rm ~~for~~} 
\pi = (1,v,s,c)
{\rm ~~~~of~~}\SO(2N)_1 .
}

\bigskip
\noindent\undertext{Selection Rules}

The branching function $ b^{\la,\pi}_{\lo,\lt,q} (\tau) $
will be non-vanishing if and only if
$(\lo,\lt,q)$ is contained in $(\la,\pi)$ or its descendants. 
This translates into a constraint on the conjugacy classes of
the representations, 
which is embodied in the following two selection rules 
\refs{\GepnerPL,\Lerche}:
\eqn\DefAlpha{\eqalign{
q & = -m r_\la+ (m+n) r_\lo + m (m+n) (a + \half n \eps) \mod mn(m+n)(m+n+k)
\cr
q & =  n r_\la- (m+n) r_\lt + n (m+n) (b + \half m \eps) \mod mn(m+n)(m+n+k)
}}
where $r_\lla$ denotes the number of boxes 
in the Young tableau corresponding to $\lla \in \SU(N)_K $
(equivalently, $r_\lla = \sum_{i=1}^{N-1}  i a_i$),
$a$ and $b$ are integers 
defined modulo $n(m+n+k)$ and $m(m+n+k)$ respectively,
and
$\eps = 0$ or 1 if 
$\pi$ belongs to the NS or R sector respectively.
Eliminating $q$ between these equations
yields a constraint between $a$ and $b$:
\eqn\Constraint{
m a - n b = r_\la - r_\lo - r_\lt \mod mn(m+n+k).
}
If $(m,n)$, the greatest common divisor of $m$ and $n$,
is greater than 1,
then eq.~\Constraint~also constrains which representations
$\la$, $\lo$, and $\lt$ are allowed: 
$r_\la - r_\lo - r_\lt $ must be a multiple of $(m,n)$.

\bigskip
\noindent\undertext{Field identification}

Not all $\La$ correspond to distinct primary fields of the coset model.
Consider the two operations \refs{\GepnerPL,\Lerche}:
\eqn\Ident{\eqalign{
\Sig_1(\La) 
&
= \pmatrix{  \si(\la) & & \si^n(\pi) \cr \si(\lo) &\lt& q + n(m+n+k) },
\cr
\Sig_2(\La) 
&
= \pmatrix{  \si(\la) & & \si^m(\pi) \cr \lo &\si(\lt)& q - m(m+n+k) },
}}
in which $\si$ is related 
to a symmetry of the extended Dynkin diagram 
of the Kac-Moody algebra, and is defined as follows:

Acting on an $\SUNK$ representation $\lla$, 
the operation $\si$ rotates the extended Dynkin indices:
$a_i (\si(\lla)) =  a_{i-1} (\lla)$,
where $a_{i+N} \equiv  a_{i}$.
Alternatively,
$\si(\lla)$ results from tensoring $\lla$ with 
the representation whose Young tableau consists of a single row of width $K$ 
(a ``cominimal'' representation \Fuchs~or simple current \SYExt).
Hence, the tableau representing 
$\si(\lla)$ is obtained from that of $\lla$
by adding a single row of $K$ boxes to the top, 
and $ r_{\si(\lla)} = r_\lla + K$ mod $N$.
We call $\lla$ and $\si(\lla)$ ``cominimally equivalent,''
and the set of representations 
$\lla$, $\si(\lla), \ldots, \si^{N-1}(\lla)$
constitute a cominimal equivalence class \Mlawer, or 
simple current orbit \SYExt.

Acting on the SO$(2N)_1$ representation $\pi$,
the operation $\si$ exchanges the extended Dynkin indices
$a_0 \leftrightarrow a_1$ and $a_{N-1} \leftrightarrow a_N$.
(This is the product of the operations $\si$ and $\varepsilon$ 
defined in  ref.~\Mlawer).
Hence $\si(1)=v$, $\si(v)=1$, $\si(s)=c$, and $\si(c)=s$.
The operation $\si$ keeps  $\pi$ within the NS or R sector.

The conformal weights and modular transformation matrices
transform under $\si$ as follows\foot{All expressions for the
$S$ matrices in ref.~\Mlawer~should be complex conjugated to agree
with those given in this paper. 
See the footnote on p.~868 of that reference.}:
\eqn\SuniOrthoComin{\eqalign{
h^{N,K}_{\si(\lla)} 
&
= h^{N,K}_\lla + {K(N-1)\over 2N}  - { r_\lla \over N}, ~~~~~~~
S^{N,K}_{\si(\lla) \lamp} = \e^{ 2\pi i r_\lamp/N} S^{N,K}_{\lla \lamp}
{\rm ~~~~~for~~SU}(N)_K
\cr
h_{\si(\pi)} 
&
= h_\pi    + \half (\eps -1) \mod \ZZ,   ~~~~~~~~~
S_{\si(\pi  ) \piep } = (-)^{\epsp}  S_{\pi   \piep }
{\rm ~~~~~~~~~~~for~~SO}(2N)_1
}}
where $\eps$ ($\epsp$) is 0 or 1 according to whether 
$ \pi$ ($\piep$) is in the NS or R sector respectively.

If $\La$ obeys the selection rules \DefAlpha,
then $\Sig_1(\La)$ and $\Sig_2(\La)$ do so as well.
Using eqs.~\CosetH~and \SuniOrthoComin, 
one may check that
$ h_{\Sig_1(\La) } = h_{\Sig_2(\La) } = h_{         \La  } $
modulo integers for $\La$ obeying eq.~\DefAlpha.
The superconformal U(1) charge \ScflCharge~is also invariant 
(mod $2\ZZ$) under $\Sig_1$ and $\Sig_2$.
Finally, from eq.~\CosetS, 
it follows that the modular transformation matrices
are invariant:
$ S_{\Sig_1(\La) \Lap} =  S_{\Sig_2(\La) \Lap} = S_{\La \Lap}$. 
This implies that the determinant of $S$ vanishes,
since it has identical columns,
and so cannot satisfy the modular group relation $S^4 = 1$.

To resolve this problem \refs{\Moore-\Lerche},
one should identify $\Sig_1(\La)$ and $\Sig_2(\La)$ with $\La$,
rather than considering them to be distinct primary fields.
The primary fields of the coset model are then equivalence
classes of $\La$, each of which contains $mn(m+n)$ elements,
except when $m$, $n$, and $k$ have a common divisor, in which
case some of the equivalence classes -- the ``fixed points'' --
are smaller.\foot{We will deal with this situation in section 6.}
When $m$, $n$, and $k$ have no common divisor,
the number of primary fields of $\Gmnk$ is given by
\eqn\NumPrim{
N_{m,n,k} =  4  \pmatrix{ m+n+k-1 \cr k}
		\pmatrix{ m+n+k-1 \cr n+k}
		\pmatrix{ m+n+k-1 \cr m+k}
                {m+n+k \over mn(m+n)},
}
which is invariant under any permutation of $m$, $n$, and
$k$.\foot{We thank M. Crescimanno for this observation.}
The ``field identification'' rescales 
the modular transformation matrix \SYCoset,
giving rise to the factor $mn(m+n)$ in eq.~\CosetS.

The connection between selection rules
and field identification is most easily understood via the
formalism of identification currents \refs{\SYCoset,\Schellekens}.
One first finds the identification currents of the coset model,
which are  simple currents with respect to which
fields that obey the selection rules \DefAlpha~have
vanishing monodromy charge.
Then one identifies any two fields related by an identification current.
That is, primary fields of the coset model 
correspond to orbits of the identification currents.
For $\Gmnk$, 
the identification  currents are $\Sig_1 (I)$ and $\Sig_2 (I)$ \Schellekens,
where $I$ denotes the identity $\pmatrix{ 1 & & 1 \cr 1  & 1  & 0  }$.

The field identifications \Ident~allow us 
some freedom in choosing 
which set of representations 
$\la$, $\lo$, $\lt$, $\pi$, and $q$
to use to describe a given primary field.
For example, we may use $\Sig_1$ 
to choose $\lo$ to be any element in its 
cominimal equivalence class.
Then we can use $\Sig_2$ to choose $\la$ to be any element
in its class, without affecting the prior choice of $\lo$.
We will choose both $\la$ and $\lo$ to be ``cominimally reduced,'' 
where a representation $\lla$ of $\SUNK$  is said to be 
cominimally reduced if its tableau has fewer than $K$ boxes 
in its first row, or  equivalently, if $a_0(\lla) \neq 0$.
(Since $\sum_{i=0}^{N-1} a_i(\lla) = K$, 
at least one of the extended Dynkin indices must be non-zero 
and so can be rotated into the zeroth position.)

Having fixed $\la$ and $\lo$,
we still have some freedom to shift $\lt$.
Acting with $\Sig_2^{(m+n)x}$ 
will leave $\la$ invariant for any integer $x$,
since $\si^{m+n}(\la) = \la$.
Since $\si^{(m+n)x} (\lt) = \si^{(m+n)x + n \xp } (\lt) $,
we can shift $\lt$ by $(m,n)$ units along its orbit
by choosing $x$ to satisfy
$(m+n)x + n \xp  = (m,n)$,
that is,
$mx/(m,n) = 1$ mod $n/(m,n)$,
which is guaranteed to have a solution between $0$ and $n/(m,n)$.
This gives the identification
\eqn\Residual{
\pmatrix{  \la & & \pi \cr \lo & \lt & q }
\approx \pmatrix{      \la & & \si^{m(m+n)x} (\pi) \cr
                       \lo & \si^{(m,n)}(\lt) & q - m(m+n)(m+n+k)x}.
}

With $\la, \lo$, and $\lt$ fixed,
there will generically be four possibilities for $\pi$
and $(m+n+k)$ allowed, inequivalent values of $q$,
corresponding to $4(m+n+k)$ distinct primary fields, 
as we will now show. 
If $(m,n) = 1$,
the inequivalent values of $a$, defined in eq.~\DefAlpha,
lie between 0 and $n(m+n+k)$,
but the constraint \Constraint~allows only
$(m+n+k)$ of these.
If $(m,n) \neq 1$, 
then acting with $\Sig_1^{m(m+n)/(m,n)}$
takes $\La$ to
$\pmatrix{ \la & & \pi \cr \lo & \lt & q + mn(m+n)(m+n+k)/(m,n) }$, 
so we identify $q$'s differing by $mn(m+n)(m+n+k)/(m,n)$.
There are then $n(m+n+k)/(m,n)$ 
inequivalent values of $a$, but
the constraint \Constraint~reduces this by a factor of $n/(m,n)$,
so again
there are (generically) $(m+n+k)$ allowed, inequivalent values of $q$.
This will not be true, however, if there are short orbits.

\bigskip
\noindent\undertext{Short orbits of simple currents}

When $(N,K) \neq 1$,
it can happen that the fields
$\lla$, $\si (\lla), \ldots, \si^{N-1}(\lla)$ 
are not all distinct but that
$\si^{N/d} (\lla) = \lla$ for some divisor $d$ of $N$.
We then say that $\lla$ is in a ``short orbit'' of length $N/d$.
The Dynkin indices of $\lla$  repeat in $d$ groups of  $N/d$:
$(a_0, a_1, \ldots, a_{N/d -1}, a_0, a_1, \dots, a_{N/d-1}, \ldots)$,
where $\sum_{i=0}^{N/d-1} a_i = K/d$,
so in fact $d$ must divide $(N,K)$.

If one or more of the representations $\la$, $\lo$, or $\lt$
is in a short orbit, then there will be fewer than 
$(m+n+k)$ inequivalent, allowed values of $q$ for fixed
$\la, \lo, \lt$, and $\pi$.
For example, 
if $\lo$ is in a short orbit of length $m/d$,
then acting with $\Sig_1^{m(m+n)/d}$
takes $\La$ to
$\pmatrix{ \la & & \si^{mn(m+n)/d} (\pi)\cr 
           \lo & \lt & q + mn(m+n)(m+n+k)/d}$, 
forcing us to identify $q$'s that differ by $mn(m+n)(m+n+k)/d$
(when $mn(m+n)/d$ is even).
There will then be $4(m+n+k)/d$ distinct primary fields
for fixed $\la$, $\lo$, and $\lt$.
When $mn(m+n)/d$ is odd, the identification interval for $q$
is doubled, but in that case $\pi$ and $\si (\pi)$ 
do not correspond to distinct primary fields, 
so the number of primary fields remains $4(m+n+k)/d$.

\newsec{Level-rank duality between $\SUNK$ and $\SUKN$}

In this section, 
we briefly review the results of level-rank duality between
the $\SUNK$ and $\SUKN$ WZW models.
Level-rank duality denotes a correspondence, 
not necessarily an equivalence, 
between various quantities of the two models.
First, it is a simple algebraic fact that the sum of central charges 
of dual theories obeys
\eqn\CDuality{
c^{N,K}  + c^{K,N} = NK - 1.
}
Second, if $\lla$ denotes an integrable representation of $\SUNK$,
then $\llat$, 
defined by exchanging the rows and columns of 
the Young tableau corresponding to $\lla$, 
is an integrable representation of $\SUKN$.
If $\lla$ is a cominimally-reduced representation, 
then the conformal weights of $\lla$ and $\llat$ 
satisfy \refs{\LevelRank, \ABI}
\eqn\HDuality{
h^{N,K}_\lla + h^{K,N}_\llat  
={r_\lla\over 2} \left( 1 - {r_\lla\over NK} \right).
}
If $\lla$ is not cominimally reduced, 
then the sum differs from this by a known \Mlawer~integer.
Similarly, the modular transformation matrices of dual theores are
related by\foot{See footnote 3.} \refs{\Kuniba-\Mlawer}
\eqn\SDuality{
S^{N,K}_{\lla \lamp} 
= \sqrt{K\over N} ~\e^{ 2\pi i r_\lla r_\lamp / NK}
\left( S^{K,N}_{\llat \lamtp} \right)^*
}
which by virtue of Verlinde's formula leads to 
an equality of fusion coefficients
of $\SUNK$ and $\SUKN$ \refs{\Kuniba, \Mlawer}
\eqn\Fusion{
N_{\lla \lamp}^{~~~\lla^{\prime\prime}}=
N_{\llat \lamtp}^{~~~\si^\Delta(\llat^{\prime\prime})},
}
where $\Delta = (r_\lla + r_\lamp - r_{\lla^{\prime\prime}})/N \in \ZZ$.

The number of primary fields of $\SUNK$,
namely $\left( N+K-1 \atop K \right)$,
is not invariant under $N \leftrightarrow K$, 
hence the transpose map 
$\lla \to \llat$ between primary fields of $\SUNK$ and $\SUKN$ 
cannot be  one-to-one.
Indeed, two cominimally equivalent representations will (often)
transpose to the same dual representation.
The transpose map, however,  {\it is} one-to-one between cominimal
equivalence classes, or simple current orbits.
Moreover, the sizes of the orbits are correlated under this map:
if $\lla$ is in a short orbit of length $N/d$,
then $\llat$ is in a short orbit of length $K/d$.
(We will demonstrate this in section 6.)
Letting $n_{\rm orbits}$ be the number of orbits
and $d_i$ the divisor of the $i^{\rm th}$ orbit, 
we can write the number of primary fields of $\SUNK$ and $\SUKN$ as 
$\sum_{i=1}^{n_{\rm orbits}} \left( N / d_i \right)$ and
$ \sum_{i=1}^{n_{\rm orbits}} \left( K / d_i \right)$
respectively.
The ratio of these is manifestly $N/K$,
consistent with the expression above.

\newsec{ Map between primary fields of $\Gmnk$ and $\Gknm$}

In this section, we provide strong evidence for the equivalence 
of the coset models
\eqn\GMNK{
\Gmnk = { \SU(m+n)_k \times \SO (2mn)_1 \
\over \SU(m)_{n+k} \times \SU(n)_{m+k} \times \U_{mn(m+n)(m+n+k)}} 
}
and
\eqn\GKNM{
\Gknm = { \SU(k+n)_m \times \SO (2kn)_1 \
\over \SU(k)_{n+m} \times \SU(n)_{m+k} \times \U_{kn(k+n)(m+n+k)}} 
}
by establishing a one-to-one map between the primary fields of the
theories, 
and showing the equivalence of their conformal weights (mod $\ZZ$),
their superconformal U(1) charges (mod $2\ZZ$),
and their modular transformation matrices (and hence fusion rules).

The difference of the central charges of these two cosets 
\eqn\CDifference{
c^{m,n,k} - c^{k,n,m}
 = (c^{m+n,k} + c^{k,m+n}) - (c^{m,n+k} + c^{n+k,m}) + (mn - kn) = 0
}
vanishes as a consequence of the level-rank relation \CDuality.
To specify a one-to-one map between the primary fields of $\Gmnk$
and $\Gknm$,
we will define a map between the multi-indices
\eqn\PrimaryMap{
\La = \pmatrix{ \la & & \pi \cr \lo & \lt & q  } ~~\to~~
\Lat = \pmatrix{ \mue & & \rho \cr \mo & \mt & \qt  }
}
up to field identifications.
The selection rules \DefAlpha~for $\Gknm$ require
\eqn\DefAlphaTilde{\eqalign{
\qt &= -k r_\mue + (k+n) r_\mo 
       + k (k+n)( \alt+\half n \epst) \mod kn(k+n)(m+n+k)\cr
\qt & =  n r_\mue - (k+n) r_\mt 
        + n (k+n) (\bet+\half k \epst) \mod kn(k+n)(m+n+k)
}}
where $\alt$ and $\bet$ are integers, and
$\epst=0 $ or 1  for $\rho  \in$ NS or R respectively.
Combining the two selection rules yields the constraint
\eqn\DualConstraint{
k \alt - n \bet = r_\mue - r_\mo - r_\mt \mod kn(m+n+k),
}
which implies that $ r_\mue - r_\mo - r_\mt $ must be 
a multiple of $(k,n)$.

\bigskip
\noindent\undertext{Determining the map}

Since the group factors $\SU(k+n)_m$ and $\SU(k)_{m+n}$ in $\Gknm$
are level-rank duals of factors in $\Gmnk$, 
the natural map
between corresponding representations is Young tableau transposition
\refs{\LevelRank-\Nakanishi, \FSAnn}: 
\eqn\TransposeMap{\eqalign{
\mue &= \lot,\cr
\mo  &= \lat. \cr
}}
That the level-rank duality map is only well-defined
between cominimal equivalence classes 
dovetails with the fact that 
the primary fields of the coset theory 
are defined by $\La$ only up to field identification.
To be precise about the map,
we first use the field identifications \Ident~to ensure
that both $\la$ and $\lo$ are cominimally reduced.

Both $\lt$ and $\mt$ are representations of $\SU (n)_{m+k}$,
but the naive guess 
\eqn\Naive{
\mt = \lt~~~~~~{\rm (incorrect)}
}
is incorrect.
To understand why this is so,
consider the chain of mappings
\eqn\Chain{
\Gmnk \to \Gknm \to G(k,m,n) \to G(n,m,k).
}
Applying the maps \TransposeMap~and \Naive~successively, we
would arrive at
\eqn\WrongDeduction{
\pmatrix{ \la &  & - \cr \lo & \lt & - } ~~\to~~
\pmatrix{ \la &  & - \cr \lt & \lo & - } ~~~~~~{\rm (incorrect)}
}
as the map from the fields of $\Gmnk$ to $G(n,m,k)$.
That this is wrong can most easily be seen 
by noting that the selection rules \DefAlpha~are
{\it not} invariant under eq.~\WrongDeduction~together with
$m \leftrightarrow n$ 
because of a relative minus sign between them.
By considering how the Dynkin diagrams of $\SU(m)$ and $\SU(n)$ 
are embedded in the Dynkin diagram of $\SU(m+n)$,
one realizes that the exchange of $m$ and $n$ 
must be accompanied by a $\ZZ_2$ flip of each Dynkin diagram, 
which takes a representation $\lla \in \SU(N)$ to its conjugate
$\llab$, since $a_i (\llab) = a_{N-i} (\lla)$.
Thus, under $m \leftrightarrow n$, 
the correct mapping  of representations is not eq.~\WrongDeduction~but
\eqn\RightDeduction{
\pmatrix{ \la &  & - \cr \lo & \lt & - } ~~\to ~~
\pmatrix{ \lab &  & - \cr \ltb & \lob & - } ~~~~~~{\rm (correct)}
}
which preserves the selection rules \DefAlpha,
since $r_\llab = - r_\lla$ mod $N$.

One obvious way of obtaining eq.~\RightDeduction ~from the chain
of mappings \Chain~ 
is to postulate
\eqn\StillWrong{
\mt = \ltb~~~~~~{\rm (incorrect)}.
}
This map, however, will not always be correct
as it may violate the selection rule constraint \DualConstraint.
Since $\la$ and $\lo$ are cominimally reduced, 
we have $r_\lat = r_\la$ and $r_\lot = r_\lo$,
hence 
$(r_\mue - r_\mo - r_\mt) = -(r_\la - r_\lo - r_\lt) \mod n $.
While $(r_\la - r_\lo - r_\lt )$
is necessarily a multiple of $(m,n)$ in order to satisfy
eq.~\Constraint, 
it will not necessarily be a multiple of $(k,n)$,
so the proposed map may not obey the constraint \DualConstraint.

The correct map from $\lt$ to $\mt$ is slightly more general:
\eqn\ConjugateMap{
\mt = \si^v (\ltb)~~~~~~{\rm (correct)}
}
with $v$ an integer to be specified, 
not always zero.
Since $\si$ adds a row of width $(m+k)$ to $\ltb$,
we have $r_{\si^v(\ltb)} = r_\ltb + (m+k) v $ mod $n$,
so  eq.~\DualConstraint~implies
\eqn\NewDualConstraint{
k \alt - n \bet = r_\lo - r_\la - [-r_\lt + (m+k)v ] \mod n,
} 
which means that $v$ must be chosen to satisfy
\eqn\VConstraint{
r_\la - r_\lo - r_\lt + (m+k) v = 0 \mod (k,n).
}
Our specification of $v$ below will automatically satisfy this constraint.

In defining the map from $\pi \in \SO (2mn)_1$ to $\rho \in \SO (2kn)_1$,
we assume that $\epst=\eps$, \ie,
the NS sector maps to the NS sector, 
and the R sector to the R sector.
Given this, only a two-fold choice remains,
namely whether $\pi=1$ maps to $\rho=1$ or to $\rho=v$, 
and similarly for other values of $\pi$.
(Although the groups $\SO(2mn)_1$ and $\SO(2kn)_1$
differ, we will with slight abuse of notation use
the same symbols for the conjugacy classes of each.
Thus $\rho=\pi$ means $1 \in \SO(2mn)_1$ maps to $1 \in \SO(2kn)_1$
and so forth.) 
We thus write this map
\eqn\OrthoMap{
\rho = \si^u (\pi)
}
where the integer $u$ is defined modulo 2.
Thus, the map between $\La$ and $\Lat$ is 
\eqn\Map{
\pmatrix{ \la & & \pi \cr \lo & \lt & q  } ~~ \to ~~
\pmatrix{ \lot & & \si^u(\pi)  \cr \lat & \si^v (\ltb) & \qt  }
}    
where $v$, $u$, and $\qt$ have yet to be specified. 

As discussed in section 2,
for fixed $\la$, $\lo$, and $\lt$,
there will generically be $4(m+n+k)$ distinct primary fields of $\Gmnk$, 
labelled by $\pi$ and $q$.
Likewise,
for fixed $\mue$, $\mo$, and $\mt$,
$\Gknm$ will have $4(m+n+k)$ primary fields,
so it is reasonable to seek a one-to-one map.
As we saw in the last paragraph of section 2, however,
if one or more of $\la$, $\lo$, or $\lt$ belong to short orbits,
the number of distinct primary fields in $\Gmnk$ is fewer,
appearing to  endanger the one-to-one correspondence.
In fact, the one-to-one correspondence is preserved:
if for example $\la$ is in a short orbit of length $(m+n)/d$,
then $\lat$ is in a short orbit of length $k/d$
(as will be shown in section 6), 
so the number of distinct primary fields of $\Gmnk$ and $\Gknm$ 
is reduced by the same factor. 

To further determine the  map \Map, we use the
invariance of the superconformal U(1) charges \ScflCharge~of 
corresponding primary fields,
$Q_\La = \Qt_\Lat$, which  implies 
\eqn\ScflEquality{
\Pie_{2kn}(\rho)  - \Pie_{2mn} (\pi) 
= {\qt - q \over m+n+k}    \mod 2.
}                  
Eqs.~\DefSigma~and \OrthoMap~imply
\eqn\SigmaDiff{
\Pie_{2kn}(\rho)  - \Pie_{2mn} (\pi) 
=  u + \half (k-m)n \eps \mod 2.
}
Likewise, the selection rules~\DefAlpha~and \DefAlphaTilde~yield
\eqn\QDiff{
      {\qt - q \over m+n+k} 
= r_\la - r_\lo 
+ (k-m) (a + \half n \eps) 
+ k(k+n) { (\alt - a) \over m+n+k} 
\mod \cases{n, & $n$ even, \cr
            2n, & $n$ odd. \cr}
}                  
Combining eqs.~\ScflEquality, \SigmaDiff~and \QDiff, we learn that
$k(k+n) (\alt-a) $ must be divisible by $m+n+k$,
but since $k(k+n)$ is not generically divisible by $m+n+k$,
we will assume that $\alt - a$ is, so that
\eqn\DefS{
\alt = a +  (m+n+k)  s
}
for some integer $s$.
The relation \DefS~between $a$ and $\alt$ then determines $u$:
\eqn\DefU{
u =  r_\la - r_\lo +  (k-m) a  + k (k+n) s \mod 2.
}
To find $s$, we add the constraints \Constraint~and \NewDualConstraint~
and use eq.~\DefS~to find
\eqn\SumConstraint{
(m+k)(a + k s + v ) = 0 \mod n.
}
If $(m+k,n)=1$,     it follows that 
$ a + k s + v $ must be divisible by $n$.
We make the plausible assumption that this remains
true even when $(m+k,n) > 1$:
\eqn\Assumption{
a + k s + v   = 0 \mod n
}
which  requires that
\eqn\DefV{
v = -a \mod (k,n).
}
Equation \DefV~will serve as our definition of $ v$.
To see that this satisfies the constraint \VConstraint,
observe that eq. \Assumption~together with eq.~\Constraint~ implies
\eqn\Consequence{
mv + m k s + r_\la - r_\lo  -r_\lt = 0 \mod n,
}
hence eq.~\VConstraint~follows.
The mod $(k,n)$ ambiguity in the definition \DefV~ of $v$
precisely corresponds to the ambiguity due to field identification,
as we will see below.
When $(k,n)=1$, one is always free to choose $v = 0$.

Having chosen some $v$ satisfying eq.~\DefV,   
we set the right hand side of eq.~\Assumption~ to $tn$, for
some integer $t$,
whence
\eqn\DefT{
-{k\over (k,n)} s + {n\over (k,n)} t 
= {a + v \over (k,n)}.
}
Since $k/(k,n)$ and $n/(k,n)$ are relatively prime,
this equation has a unique solution $s$ modulo $n/(k,n)$.
By eqs.~\DefS~and \DefAlphaTilde,
$s$ determines $\qt$ modulo $kn(k+n)(m+n+k)/(k,n)$,
which from our discussion in section 2 is precisely
the expected ambiguity due to field identification.

Given this solution $s$
(chosen, say,  to lie in the range $0, \ldots, {n \over (k,n)} - 1$),
we may now rewrite 
\eqn\AlphaResult{\eqalign{
a &= -   k s + n t - v
\cr
\alt &= (m+n) s + n t - v
}}
and from eq.~\DefU,
\eqn\UResult{
u = r_\la - r_\lo +  k (m+n) s  + (k-m) (  n t - v ),
}
all in terms of $s$, $t$, and $v$.

Let us now reconsider the mod $(k,n)$ ambiguity 
in the definition of $v$ in eq.~\DefV.
Suppose instead of $v$ we chose 
\eqn\DefNewV{
\vp = v + (k,n).
}
Then using eqs.~\DefT~and \DefU, we have
\eqn\DefNewS{
s^\prime = s - \xt, ~~~~~\uup = u - k(k+n)\xt
}
where $ k\xt/(k,n) = 1 \mod n/(k,n)$.
With this choice of $v$, the map from $\Gmnk$ to $\Gknm$ becomes
\eqn\MapAmbiguity{
\pmatrix{ \la & & \pi \cr \lo & \lt & q  } \rightarrow
\pmatrix{ \lot & & \si^{u - k(k+n) \xt } (\pi)  \cr
          \lat & \si^{v + (k,n)} (\ltb) & \qt-k (k+n)(m+n+k) \xt  }.
}    
The difference between eqs.~\Map~and \MapAmbiguity,  however, is 
just the field equivalence \Residual.

\vfil\break
\noindent\undertext{Equivalence of conformal weights 
and modular transformation matrices}

Having determined the map between primary fields
of $\Gmnk$ and $\Gknm$, we now proceed to establish
the equivalence of their conformal weights and modular 
transformation matrices.
The conformal weights \CosetH~
of $\La \in \Gmnk$ and $\Lat \in \Gknm$ differ by 
\eqn\HDiff{\eqalign{
h^{m,n,k}_\La - h^{k,n,m}_\Lat ~=&~
 (h^{m+n,k}_\la + h^{k,m+n}_\lat)
 - (h^{m,n+k}_\lo + h^{n+k,m}_\lot) \cr
& + (h^{n,m+k}_{\si^v (\ltb)}  - h^{n,m+k}_\lt) 
 + (h^{2mn}_\pi - h^{2kn}_{\si^u (\pi)} )
 + (h_\qt - h_q) \mod \ZZ. \cr
} }
The first two pieces of this are given by the level-rank relation \HDuality,
\eqn\HNoughtDuality{\eqalign{
h^{m+n,k}_\la + h^{k,m+n}_\lat 
&=    {r_\la\over 2} \left[ 1 - {r_\la\over k(m+n)} \right],
\cr
h^{m,n+k}_\lo + h^{n+k,m}_\lot
&={r_\lo \over 2} \left[ 1 - {r_\lo\over m(n+k)} \right].
}}
We also need \Mlawer
\eqn\HTwoDiff{
  h^{n,m+k}_{\si^v (\ltb)} ~=~ 
h^{n,m+k}_{\si^{-v} (\lt)} ~=~ 
h^{n,m+k}_\lt 
+ { (m+k) v (n-v) \over 2n}  + { v r_\lt \over n} \mod \ZZ
}
together with \WZWH
\eqn\HOrthoDiff{
h^{2mn}_\pi - h^{2kn}_{\si^u(\pi)} 
    = \cases{ \vphantom{a_1\over a_1}  \half u \mod \ZZ, & $\pi \in$ NS \cr
              \vphantom{a_1\over a_1}  \eighth n (m-k), & $\pi \in$ R.  \cr}
}
The difference of U(1) weights is
\eqn\HUniDiff{\eqalign{ 
h_\qt - h_q 
 ~=& ~  {\qt^2 \over 2kn(k+n)(m+n+k)}
  -  {q^2 \over 2mn(m+n)(m+n+k)}    \cr
 ~=& ~   {r_\la^2 \over 2k(m+n)} 
  - {r_\lo^2 \over 2m(k+n)}    
 - ~{v\over n} \left[ r_\la - r_\lo + \half (m-k) v  + m k s \right] 
 + \half  \eps \left( r_\la - r_\lo \right)  \cr
 &\vphantom{ {r_\la^2 \over 2k(m+n)}}
   + \half  \eps (m-k) v 
  + \half n (k-m) \left(t + \half \eps \right)^2   
  + \half k (m+n) s (s + \eps)  \mod \ZZ \cr
}}
via eqs.~\DefAlpha, \DefAlphaTilde, and \AlphaResult.
Putting all these together, and using eq.~\Consequence,    
it is straightforward to see that 
$h^{m,n,k}_\La - h^{k,n,m}_\Lat = 0$ mod $\ZZ$
in the R ($\eps=1$) sector.
In the NS ($\eps=0$) sector,
\eqn\Neveu{
h^{m,n,k}_\La - h^{k,n,m}_\Lat
  = \half \left[r_\la - r_\lo + (m-k) v  +   n(k-m) t^2
+ k (m+n) s^2 + u \right] \mod \ZZ.
}
Using the expression \UResult~for $u$, this reduces to
\eqn\Schwarz{
h^{m,n,k}_\La - h^{k,n,m}_\Lat
=  \half \left[ n (k-m) t (t+1)
          + k (m+n) s (s+1)  \right]   \mod \ZZ
}
which vanishes mod $\ZZ$.
Thus we have shown that in general
\eqn\HEquality{
h^{m,n,k}_\La = h^{k,n,m}_\Lat \mod \ZZ.
}

Next, consider the ratio of modular transformation matrices \CosetS~for
$\Gmnk$ and $\Gknm$:
\eqn\SRatio{
{ S^{m,n,k}_{\La \Lap} \over  S^{k,n,m}_{\Lat \Latp} }
= {m(m+n)\over k(k+n)}
{ S_{\la \lap} \over  S^*_{\lat \latp} }
{ S^*_{\lo \lop} \over  S_{\lot \lotp} }
{ S^*_{\lt \ltp} \over  S^*_{\si^v(\ltb) \si^\vp(\ltbp)} }
{ S_{\pi \piep} \over  S_{\si^u(\pi) \si^\uup(\piep)} }
{ S^*_{q \qp} \over  S^*_{\qt  \qtp} }.
}
Again we need the level-rank relations \SDuality
\eqn\SuniSRatio{\eqalign{
{ S_{\la \lap} \over  S^*_{\lat \latp} }
&
= \sqrt{k\over m+n} \exp \left[ {2\pi i r_\la r_\lap \over k(m+n)} \right],
\cr
{ S^*_{\lo \lop} \over  S_{\lot \lotp} }
&
= \sqrt{n+k\over m} \exp \left[-~ {2\pi i r_\lo r_\lop \over m(n+k)} \right],
}}
together with\foot{See footnote 3.} \Mlawer
\eqn\SuniSTwoRatio{
{ S^*_{\lt \ltp} \over  S^*_{\si^v(\ltb) \si^\vp(\ltbp)} }
= \exp \left( {2\pi i \over n} 
\left[  (m+k)v \vp - v r_\ltp - \vp r_\lt \right] \right)
}
and \OrthoS
\eqn\OrthoSRatio{
{ S_{\pi \piep} \over  S_{\si^u(\pi) \si^\uup(\piep)} }
= \cases { 1 & $\pi , \piep \in$ NS, \cr
           (-1)^u & $\pi \in$ NS, $\piep \in$ R,\cr 
           (-1)^\uup &$ \pi \in$ R, $\piep \in$ NS,\cr 
           (-1)^{u+\uup}i^{n(k-m)} & $\pi , \piep \in$ R.\cr }
}
The ratio of the U(1) modular transformation matrices
\eqn\UniSRatio{ \eqalign{
{ S^*_{q \qp} \over  S^*_{\qt  \qtp} }
& = \sqrt{ k(k+n) \over m(m+n) }
\exp \left( {2 \pi i\over m+n+k}
\left[   {q \qp \over mn(m+n)} - {\qt \qtp \over kn(k+n)} \right] 
\right) \cr
& = \sqrt{ k(k+n) \over m(m+n) }
\exp \left( 2 \pi i 
\left[ - ~{r_\la r_\mue \over k(m+n)} + {r_\lo r_\mo \over m(k+n)}
+ {v r_\ltp + \vp r_\lt \over n}  \right. \right. \cr 
& ~~~~~~~~~~~~~~~~~~~~~~~~~~~~~~~~~\left. \left.-~ {(k+m) v \vp \over n}
 - {\eps \uup + \epsp u \over 2} + {\eps \epsp n (m-k) \over 4}
\right] \right) \cr
} }
is obtained using eqs.~\Consequence~and \UResult.
The factors in the product \SRatio~exactly cancel,
establishing the equivalence of the modular transformation matrices
\eqn\SEquality{
S^{m,n,k}_{\La \Lap} =  S^{k,n,m}_{\Lat \Latp} .
}
The equivalence of the fusion rules then automatically follows from
Verlinde's formula.

\newsec{ Map between the chiral rings of G$(m,1,k)$ and G$(k,1,m)$}

In this section, we describe the map between the chiral rings of
$G(m,n,k)$ and $G(k,n,m)$ when $n=1$.\foot{In this case,
the Kazama-Suzuki model is a Landau-Ginzburg theory,
but not necessarily otherwise \Lerche.}
The chiral ring is composed of chiral primary fields,
those that saturate the bound $h_\La \ge \half \left| Q_\La\right|$.
For $G(m,1,k)$ these have a simple characterization \GepnerPL
\eqn\ChiralPrimary{
\La^{\rm chiral} = \pmatrix{ \la    & & 1  \cr 
                             \lo    &-& q \cr} 
}
where, if the Dynkin indices of $\la \in \SU(m+1)_k$ are
$(a_1, \ldots, a_m)$, then
the Dynkin indices of  $\lo \in \SU(m)_{k+1} $ are 
$(a_1, \ldots, a_{m-1})$, and 
$q = r_\la$.
The chiral primaries are in one-to-one correspondence with the primary
fields of $\SU(m+1)_k$, which number $\left( m+k \atop k \right)$,
and their conformal weights and superconformal U(1) charges 
are proportional to the number of boxes of the corresponding tableau,
$h_\La = -~ \half Q_\La =  r_\la/[2(m+1+k)] $ mod $\ZZ$.
In fact the Poincar\'e polynomial \Lerche~for $G(m,1,k)$ just
counts the number of $\SU(m+1)_k$ representations 
graded by the number of boxes in their tableaux.

The number of chiral primaries is manifestly invariant under 
$m \leftrightarrow k$.
What is the relation between the chiral primary \ChiralPrimary~of
$G(m,1,k)$ and the corresponding chiral primary
\eqn\DualChiral{
\Lat^{\rm chiral} = \pmatrix{ \mue &  & 1  \cr 
                               \mo &- & \qt\cr} 
}
of $G(k,1,m)$?
The answer is that $\mue \in \SU(k+1)_m$ is given by 
$\lat$, the tableau transpose of $\la \in \SU(m+1)_k$,
and $\qt= q$, since tableau transposition preserves the number of boxes.
Note that tableau transposition generates a one-to-one map between
the primary fields of $\SU(m+1)_k$ and $\SU(k+1)_m$.\foot{See
conjecture 1 of ref.~\Zuber.}
This is in contrast to the usual level-rank duality between
$\SUNK$ and $\SUKN$  in which tableau transposition only generates
a correspondence between cominimal equivalence classes.

At first sight, this map between chiral primaries seems different from
the map between primary fields
prescribed in the previous section, 
in which $\mue = \lot$ and $ \mo  = \lat$, but
we will show that the two maps are equivalent.
First we consider the case in which $\la$, and therefore $\lo$,
are cominimally reduced.
Then, by the prescription given in section 4, 
the chiral primary \ChiralPrimary~maps to
\eqn\MappedChiral{
\Lat^{\rm chiral} = \pmatrix{ \lot   &  & \si^{k a_m}(1)   \cr 
                              \lat   & -& q + k(m+1+k)a_m \cr} .
}
where $a_m$ is the last Dynkin index of $\la$.
By acting  $k(k+1-a_m)$ times with $\Sig_1$ on this field, we obtain
\eqn\ShiftedChiral{
\Lat^{\rm chiral} = \pmatrix{ \si^{a_m} (\lot)   &  & 1   \cr 
                              \lat               & -& q  \cr} 
}
but since the tableau for $\lo$ is the same as the tableau for $\la$
with $a_m$ columns of $m$ boxes preprended to it, it follows that
$\si^{a_m} (\lot)$ is simply $\lat$,
so the field \ShiftedChiral~is just eq.~\DualChiral.

Next suppose $\la$ is not cominimally reduced, but has $\ta_k$
rows of boxes of width $k$ at the top.
To implement the map to $G(k,1,m)$,
we first need 
to cominimally reduce $\la$ by acting $m \ta_k$ times with $\Sig_1$,
which gives
\eqn\ReducedChiral{
\La^{\rm chiral} = \pmatrix{ \si^{-\ta_k}(\la)  & & \si^{m \ta_k}(1)  \cr 
                              \lo               &-& q +m(m+1+k)\ta_k \cr} .
}
Then we map this to G$(k,1,m)$, obtaining
\eqn\TransRedChiral{
\Lat^{\rm chiral} = \pmatrix{ \lot           &  & \si^{k \ta_m}(1)     \cr 
                        \widetilde{\si^{-\ta_k}(\la)}&-& q +k(m+1+k)a_m \cr} .
}
Finally, we act $k(k+1-a_m)$ times with $\Sig_1$ to obtain
\eqn\FinalChiral{
\Lat^{\rm chiral} = \pmatrix{ \si^{a_m} (\lot)   &  & 1     \cr 
                        \widetilde{\si^{-\ta_k}(\la)}&-& q \cr} .
}
Since
$\si^{a_m} (\lot) $ is just $\lat$,
and $\widetilde{\si^{-\ta_k}(\la)}$ is just $\lat$ with $\ta_k$ columns
of height $k$ removed,
this is precisely the field \DualChiral.
Thus, the map between primary fields defined in section 4 
is equivalent to the map between chiral primaries described
below eq.~\DualChiral.

\newsec{ Fixed point resolution}

If $m, n$, and $k$ have a common divisor $p>1$,
then some of the equivalence classes of fields
$\La =\pmatrix{ \la & & \pi \cr \lo & \lt & q  } $
have fewer than $mn(m+n)$ elements.
These are called fixed-point fields,
and are those 
for which $\la$, $\lo$, and $\lt$ all belong to short orbits \Schellekens:
\eqn\ShortOrbits{
\si^{ \mh + \nh  } (\la) = \la,~~~~~             
\si^{ \mh        } (\lo) = \lo,~~~~~          
\si^{ \nh        } (\lt) = \lt,~~~~~        
}
where 
$\mh = m/p$, 
$\nh = n/p$, and 
$\kh = k/p$.
Each fixed-point equivalence class $\La$ 
actually corresponds to a set of $p$ distinct primary fields,
distinguished by the index $i$, 
\eqn\FixedPointField{       
\La^{\rm FP}_i 
= \pmatrix{ \la & & \pi& \cr \lo & \lt & q;&i },~~~~~~~i=1,\ldots p.
}
Due to this resolution of fixed points,
the characters and modular transformation matrices are modified \SYCoset~from 
the naive prescriptions \Modular~ and \CosetS.
Each row and column of the modular matrix $S$
corresponding to a fixed-point field is resolved into $p$ rows
and columns, with
\eqn\ResolvedS{
S^{\rm resolved}_{\La_i \Lap_\ip} 
 = {1\over p^2} S^{m,n,k}_{\La \Lap} E_{i\ip}
                   + \Gamma^{m,n,k}_{\La_i \Lap_\ip} 
}     
for some $  \Gamma_{\La_i \Lap_\ip} $,
where  $E_{i\ip}$ is 1 for any $i$ and $\ip$.
The conformal weights of the fields $\La_i$ are independent of $i$,
so the modular matrix $T$ is not modified.

Schellekens and Yankielowicz \SYCoset~show that the modular group
relations 
$(ST)^3 = S^2$ and $S^4 = 1$ 
obeyed by the resolved modular transformation matrices 
imply that $\Gamma$ and $T$,
restricted to the fixed-point fields, 
satisfy the same relations.
This suggests that they may be identified 
(up to a 12${}^{\rm th}$ root of unity, 
which preserves the modular group relations)
as the modular transformation
matrices $\Sh$ and $\Th$
of an auxiliary ``fixed-point'' theory \SYReview,
\eqn\DefFPMod{\eqalign{
\Gamma^{m,n,k}_{\La_i \Lap_\ip}
&  = \e^{-i \pi w/2}  \Sh^{m,n,k}_{\Lah \Lahp} P_{i \ip} \cr
T_{\La_i \La_i}
&  = \e^{ i \pi w/6}  \Th_{\Lah \Lahp}  \cr
}}     
where $P_{i\ip} = \delta_{i \ip} - (1/p) E_{i \ip}$,
$w$ is some integer,
and $\Lah$ denotes the ``projection'' of the fixed-point field $\La$
onto a field in the fixed-point theory.

For $p$ prime, 
Schellekens \Schellekens~has shown that the 
fixed-point theory corresponding to $\Gmnk$ is 
\eqn\HatCosetDef{
\Ghmnk
= { \SU(\mh+\nh)_\kh \times \SO (2mn)_1 \over
           \SU(\mh)_{\nh+\kh} \times \SU(\nh)_{\mh+\kh} 
          \times \U_{\mh \nh (\mh+\nh)(\mh+\nh+\kh)}}.
}    
Observe that $ \Ghmnk$
differs from $ G(\mh,\nh,\kh) $ 
in that the orthogonal group factor is
SO$(2mn)_1$, not SO$(2\mh \nh)_1$.
Given this fixed-point theory,
we need to determine the projection $\La \to \Lah$.
Recall that the Dynkin indices of a representation $\lla$ of $\SUNK$ 
in a short orbit repeat in groups of $\Nh$,
$(a_0, a_1, \ldots, a_{\Nh - 1}, a_0, a_1, \ldots, a_{\Nh - 1}, \ldots)$,
since $\si^{\Nh } (\lla) =\lla $. 
Here $\Nh = N/p$ and $\Kh = K/p$.
We associate \SYApp~with $\lla$ a  representation $\llah$ of 
$ \SU(\Nh)_{\Kh}$ with Dynkin indices $(a_0, a_1, \ldots, a_{\Nh-1})$.
One may then show that 
\eqn\Boxes{
r_\lla = p r_\llah + \half p (p-1) \Nh \Kh
}
and \SYApp
\eqn\SuniHH{
\left( h^{N,K}_\lla - {c^{N,K}\over 24} \right)
=
\left( h^{\Nh,\Kh }_\llah - {c^{\Nh,\Kh}\over 24} \right)
+ {NK - \Nh \Kh \over 24}.
}
Since $\la$ and $\lo$ are in short orbits,
$q$ is divisible by $p^2$,
by eqs.~\DefAlpha~and \Boxes.
Moreover, $\Sig_1^{m(m+n)/p^2}$
acting on a fixed-point field allows us to identify
$q$'s differing by $mn(m+n)(m+n+k)/p^2$.
Hence, we may rescale $q$ to 
$\qh= q/p^2$,
and regard $\qh$ as labeling a representation of
U$(1)_{\mh \nh (\mh +\nh)(\mh +\nh +\kh)}$.
In short, $\La$ is projected onto the multi-index
\eqn\HatField{              
\Lah =\pmatrix{ \lah & & \pi  \cr \loh & \lth & \qh }
}                           
belonging to $\Ghmnk$.

Using eqs.~\CosetH~and \SuniHH,
we calculate the difference between the conformal weights
of $\La \in \Gmnk$ and $\Lah \in \Ghmnk$
\eqn\CosetHH{
\left( h^{m,n,k}_\La - {c^{m,n,k} \over 24} \right)
=
\left( \hh^{m,n,k}_\Lah - {\ch^{m,n,k} \over 24} \right)
+ {\mh \nh (1-p^2) \over 12} \mod \ZZ
}
where
$\hh^{m,n,k}_\Lah $ and $\ch^{m,n,k} $ are the conformal weights
and central charge of the fixed-point theory.
This implies via eq.~\DefFPMod~that $w = \mh \nh (1-p^2)$ and thus
\eqn\DefGamma{
\Gamma^{m,n,k}_{\La_i \Lap_\ip} 
= \exp \left( 2 \pi i {\mh \nh (p^2-1) \over 4} \right) 
\Sh^{m,n,k}_{\Lah \Lahp} P_{i \ip} ~~~~~~~~~~~{\rm for~}p{\rm~prime.}
}
Although no other resolution of the fixed points is known,
no proof exists that this solution is unique.

\bigskip
\noindent\undertext{Field identification in the fixed-point theory}

The identifications \Ident~of fields in $\Gmnk$
induces an identification of fields in $ \Ghmnk$
\eqn\HatIdent{\eqalign{
\Sig_1(\Lah) 
&
= \pmatrix{  \si(\lah) & & \si^n(\pi)     \cr 
             \si(\loh) &  \lth & \qh + \nh (\mh+\nh+\kh) },
\cr
\Sig_2(\Lah) 
&
= \pmatrix{  \si(\lah) & & \si^m(\pi)               \cr 
           \loh &\si(\lth) & \qh - \mh(\mh+\nh+\kh)   }.
}}
Observe that $\si^n$, not $\si^\nh$, acts on $\pi \in \SO(2mn)_1$.
To fully define the action of $\Sig_1$ and $\Sig_2$ on
the resolved fixed-point fields $\La^{\rm FP}_i$, 
we need to specify further that
\eqn\IdentFixedPoint{ \eqalign{
\Sig_1 (i) & = \cases{ i, & prime $p > 2$ \cr
                      \si^\nh (i), & $p=2$  }
               \cr
\Sig_2 (i) & = \cases{ i, & prime $p > 2$ \cr
                      \si^\mh (i), & $p=2$  }
               \cr
}}
where $\si(1)=2$ and $\si(2)=1$.
Since
\eqn\FixedPointComin{\eqalign{
P_{\si(i) \ip} &= - P_{i \ip} \cr
E_{\si(i) \ip} &= E_{i \ip} \cr
} ~~~~~~~{\rm for~} p=2
}
the assignment \IdentFixedPoint~guarantees via eq.~\Boxes~that
$ \Gamma^{m,n,k}_{\Sig_1 (\La_i) {\Lap}_\ip} 
= \Gamma^{m,n,k}_{\Sig_2 (\La_i) {\Lap}_\ip} 
= \Gamma^{m,n,k}_{\La_i          {\Lap}_\ip}$, 
so that the resolved modular transformation matrix \ResolvedS~remains
invariant under $\Sig_1$ and $\Sig_2$,
as required for field identification.

In the case we are considering, where $m$, $n$, and $k$ have
a prime greatest common divisor $p$,
the number of primary fields in $\Gmnk$ is given by
\eqn\NumPrimFP{\eqalign{
N_{m,n,k} 
& 
=  4 \biggl[
  		\left( m+n+k-1 \atop k \right)
		\pmatrix{ m+n+k-1 \cr n+k}
		\pmatrix{ m+n+k-1 \cr m+k} \biggr. \cr
& ~~~~~~~\biggl.-	\pmatrix{ \mh+\nh+\kh-1 \cr \kh}
		\pmatrix{ \mh+\nh+\kh-1 \cr \nh+\kh}
		\pmatrix{ \mh+\nh+\kh-1 \cr \mh+\kh}
 		\biggr]
                {m+n+k \over mn(m+n)} \cr
& ~~~~~+ 4p	\pmatrix{ \mh+\nh+\kh-1 \cr \kh}
		\pmatrix{ \mh+\nh+\kh-1 \cr \nh+\kh}
		\pmatrix{ \mh+\nh+\kh-1 \cr \mh+\kh}
                {\mh+\nh+\kh \over \mh\nh(\mh+\nh)} \cr
& =  { 4 (m+n+k)! (m+n+k-1)!^2\over
       m! n! k! (m+n)! (m+k)! (n+k)! }
{}~+~ \left( p - {1\over p^2} \right)
{ 4 (\mh+\nh+\kh)! (\mh+\nh+\kh-1)!^2\over
       \mh! \nh! \kh! (\mh+\nh)! (\mh+\kh)! (\nh+\kh)! } \cr
}}
where the factor of $p$ in the third line
is the multiplicity of the resolved fixed-point fields.
The expression \NumPrimFP~is 
manifestly invariant under $k \leftrightarrow m$,
so a one-to-one map between resolved primary fields
of $\Gmnk$ and $\Gknm$ is still possible.

\bigskip
\noindent\undertext{Map between fixed-point fields}

Under level-rank duality,
short orbits of $\SUNK$ of length $N/p$
are mapped onto short orbits of $ \SU(K)_N$ of length $K/p$.
The proof of this is simple:
let $\lam$ belong to an orbit of length $N/p$,
and no shorter than $N/p$.
Project it onto $\lamh$, a representation of $ \SU(\Nh)_\Kh$.
The transpose of $\lamh$ is $\lamht$, 
a representation of $ \SU(\Kh)_\Nh$.
But $\lamht$ is equal to $\lamth$, 
the projection of $\lamt$ onto $ \SU(\Kh)_\Nh$.
Therefore $\lamt$ belongs to an orbit of $\SUKN$ no longer than $K/p$.
Reversing the argument guarantees that it also belongs to an orbit
no shorter than $K/p$.

For a fixed-point field $\La$,
the representations $\la$, $\lo$, and $\lt$ are all in 
short orbits \ShortOrbits.
The argument of the previous paragraph implies
$\si^{\kh} (\lat) = \lat$ and 
$\si^{\kh+\nh} (\lot) = \lot$.
It is also true that
$\si^{\nh}( \si^v (\ltb)) = \si^v (\ltb) $,
so $\Lat$ is a fixed-point field of $\Gknm$.
Thus, resolved fixed points of $\Gmnk$ are mapped to resolved
fixed points of $\Gknm$, 
$\La^{\rm FP}_i \to \Lat^{\rm FP}_{\tau(i)}$,
where $\tau(i)$ has not yet been specified.

The resolved modular transformation matrix for fixed points
of $\Gknm$ is
\eqn\ResolvedSDual{
S^{\rm resolved}_{\Lat_i {\Latp}_\ip } 
 = {1\over p^2} S^{k,n,m}_{\Lat \Latp} E_{i\ip}
                   + \Gamma^{k,n,m}_{\Lat_i {\Latp}_\ip } 
}     
with
\eqn\DefGammaDual{
\Gamma^{k,n,m}_{\Lat_i {\Latp}_\ip}
= \exp \left( 2 \pi i {\kh \nh (p^2-1) \over 4} \right) 
\Sh^{k,n,m}_{\Lath \Lath^\prime} 
P_{i \ip} ~~~~~~~~~~~{\rm for~}p{\rm~prime,}
}
where $ \Sh^{k,n,m}_{\Lath \Lath^\prime} $ is the modular transformation
matrix of the fixed-point theory $\Gh (k,n,m)$.
Using eq.~\Map, the field $\Lat$ is projected onto
\eqn\FPMap{
\Lath = \pmatrix{ \loth &                           & \si^u(\pi)  \cr 
                  \lath & \widehat{ \si^v (\ltb) } & \qth  }
}    
with $\qth = \qt/p^2$.  Equation \FPMap~may also be written
\eqn\NewFPMap{ 
\Lath = \pmatrix{ \loht & & \si^u(\pi)  \cr 
                   \laht & { \si^v (\lthb)} & \qth    }
}              
so the map between fields of the fixed-point theories
$\Ghmnk$ and $\Gh (k,n,m)$ is 
\eqn\Quadrangle{
\pmatrix{ \lah & & \pi  \cr \loh & \lth & \qh} ~~\to~~
\pmatrix{ \loht & & \si^u(\pi)  \cr 
\laht & { \si^v (\lthb)} & \qth}.
}              
            
\bigskip
\noindent\undertext{Equivalence of resolved modular transformation matrices}

We now show the equivalence of the resolved modular transformation
matrices \ResolvedS~and \ResolvedSDual.
The equivalence of 
$S^{m,n,k}_{\La \Lap} E_{i \ip}$ 
and 
$S^{k,n,m}_{\Lat \Latp} E_{\tau(i) \tau(\ip)}$ 
was previously established in section 4, independent of $\tau(i)$.
Consider the ratio
\eqn\GammaRatio{
{ \Gamma^{m,n,k}_{\La_i {\La^\prime}_\ip} \over 
  \Gamma^{k,n,m}_{\Lat_{\tau(i)} {\Lat^\prime}_{\tau(\ip)} }}
= \e^{ \pi i \nh (\mh - \kh) (p^2-1)/2 }
 {\mh (\mh + \nh) \over \kh (\kh + \nh)}
{ S_{\lah \lahp} \over  S^*_{\laht \lahtp} }
{ S^*_{\loh \lohp} \over  S_{\loht \lohtp} }
{ S^*_{\lth \lthp} \over  S^*_{\si^v(\lthb) \si^\vp(\lthbp)} }
{ S_{\pi \piep} \over  S_{\si^u(\pi) \si^\uup(\piep)} }
{ S^*_{\qh \qhp} \over  S^*_{\qth  \qthp} }
{ P_{i \ip} \over P_{\tau(i) \tau(\ip)} }
}
where
\eqn\SHatRatio{\eqalign{
{ S_{\lah \lahp} \over  S^*_{\laht \lahtp} }
&= \sqrt{\kh \over \mh+\nh} 
\exp \left[  {2\pi i r_\lah r_\lahp \over \kh(\mh+\nh)} \right],
\cr
{ S^*_{\loh \lohp} \over  S_{\loht \lohtp} }
&= \sqrt{\nh+\kh\over \mh} 
\exp \left[-~ {2\pi i r_\loh r_\lohp \over \mh(\nh+\kh)} \right],
\cr
{ S^*_{\lth \lthp} \over  S^*_{\si^v(\lthb) \si^\vp(\lthbp)} }
&= \exp \left( {2\pi i \over \nh} 
\left[  (\mh+\kh)v \vp - v r_\lthp  - \vp r_\lth \right]
 \right),
}}
and
\eqn\SUniHatRatio{ 
	{ S^*_{\qh \qhp} \over  S^*_{\qth  \qthp} }
 =   \sqrt{ \kh (\kh + \nh)  \over \mh (\mh+\nh) }
     \exp \left( {2 \pi i\over \mh+\nh+\kh}
\left[   {\qh \qhp \over \mh \nh (\mh+\nh) } 
       - {\qth \qthp \over \kh \nh (\kh+\nh) } \right] 
\right) 
= { S^*_{q \qp} \over  S^*_{\qt  \qtp} }.
}
The easiest way to evaluate eq.~\GammaRatio~is 
to take its ratio with eq.~\SRatio~and use eq.~\Boxes~to find
\eqn\RatioOfRatios{
{ \Gamma^{m,n,k}_{\La_i {\La^\prime}_\ip} \over 
  \Gamma^{k,n,m}_{\Lat_{\tau(i)} {\Lat^\prime}_{\tau(\ip)}} }
=
\exp \left(   \pi  i (p -1)
 \left[ r_\lah - r_\loh - (\mh + \kh) v
      + r_\lahp - r_\lohp - (\mh + \kh) \vp \right] \right)
{ P_{i \ip} \over P_{\tau(i) \tau(\ip)} }
}
for $p$ prime.
If we define
\eqn\DefTau{
\tau(i) = \cases{ i, & prime $p > 2$ \cr
                  \si^{r_\lah - r_\loh - (\mh + \kh) v}  (i), & $p=2$ \cr}
}
and use eq.~\FixedPointComin,
the ratio \RatioOfRatios~becomes unity, 
and the equivalence of the resolved modular transformation matrices 
\ResolvedS~and \ResolvedSDual~is established.
Consequently, the fusion rules between the resolved primary fields
of $\Gmnk$ and $\Gknm$ are identical.

\newsec{Concluding remarks}

We have provided strong evidence that the Kazama-Suzuki models
$\Gmnk$ and $\Gknm$ are isomorphic by constructing a one-to-one
map between their primary fields and demonstrating the equivalence
of corresponding conformal weights, superconformal U(1)  charges,
modular transformation matrices, and fusion rules.
We have shown that the equivalence 
continues to hold when $m$, $n$, and $k$ possess
a common prime divisor $p$ and the theories contain fixed points.
(We expect the equivalence to hold for any $m$, $n$, and $k$,
but the proof of this would require knowledge 
of the fixed-point theory when their greatest common divisor 
is not prime.)
Primary fields corresponding to resolved fixed points in $\Gmnk$
are mapped onto resolved fixed points of $\Gknm$.
This is simpler than in other superconformal coset models
\FSAnn~in which resolved fixed-point fields are mapped to nonfixed points
and vice versa,
and thus in which fixed-point resolution is indispensible for the map.
There is, however, some subtlety in the map between the resolved
fixed points \Quadrangle~and \DefTau~when $p=2$.

\medskip
\noindent {\bf Acknowledgements:}

We thank M. Crescimanno and H. Rhedin for useful discussions.  

\listrefs
\end